# 3-dimensional arrays of BiSrCaCuO-2212 intrinsic Josephson junctions and the zero-crossing Shapiro steps at 760 GHz

H. B. Wang[a), b)]

CREST, Japan Science & Technology Corporation (JST), Kawaguchi, Japan

P. H. Wu

Research Institute of Superconductor Electronics, University of Nanjing, Nanjing 210093, China

J. Chen[c)], K. Maeda

Research Institute of Electrical Communication, Tohoku University, Sendai 980-8577, Japan

T. Yamashita[c)]

New Industry Creation Hatchery Center, Tohoku University, Sendai, Japan

Abstract

On a slice of $Bi_2Sr_2CaCu_2O_{8+x}$ 200~300 nm thick, stacks of intrinsic Josephson junctions are patterned, connected to each other, and integrated with a bow-tie antenna and some chokes. Over an area of 150 microns by 170 microns, so far we are able to fabricate up to 256 stacks with over 11,000 junctions. Such 3-dimensional arrays of intrinsic Josephson junctions respond sensitively to irradiation at 760 GHz from a far-infrared laser, showing clear zero-crossing Shapiro steps up to 0.75 V. In addition, possible applications in quantum voltage standard etc are also discussed here. The high density of integration of Josephson junctions is also promising for many other applications.

---

[a)] Electronic mail: hbwang@riec.tohoku.ac.jp

[b)] Also at (mail address) Research Institute of Electrical Communication, Tohoku University, Sendai 980-8577, Japan

[c)] Also at CREST, Japan Science & Technology Corporation (JST), Kawaguchi, Japan

Consisting of superconducting $CuO_2$ layers and non-superconducting layers at an atom scale along a *c*-axis, a piece of high-temperature superconducting single crystal is actually a one-dimensional array of intrinsic Josephson junctions (IJJs) when *a-b* plane sizes are reduced [1, 2]. For many reasons (e.g., homogeneity of junction properties; higher impedance compared with that of a single junction; easiness of fabrication; etc) such a one-dimensional array is attractive. Although one can easily obtain hundreds or thousands of junctions in one stack, with so many junctions it can be sometimes difficult to avoid performance deviations, to have all of them phase locked to external irradiation, and to distribute the applied power evenly among them. Besides, precautions should be taken to prevent junctions or their superconductivity from being damaged by Joule heating or quasiparticle injection[3, 4]. Thus, one of the major efforts in the area has been to reduce, or to limit, the junction number in a stack in a well-controlled way. On the other hand, however, a great number of junctions are quite necessary for practical applications. For example, a few thousand or a few tens of thousand junctions in series to each other are required in quantum voltage standard to get zero-crossing Shapiro steps at sufficiently high voltages for 1 V or 10 V standard calibrations[5, 6]. With Nb-based junctions both two-dimensional (2D) and three-dimensional (3D) arrays have been reported using a stacked-junction fabrication technique[7, 8], remarkably increasing the package density. And the purpose of the work to be discussed here is to develop similar structures using high temperature single crystals so that the possibilities intrinsic junctions provide us with can be fully explored.

As reported earlier [9], using a novel double-side process we have been able to fabricate one-dimensional (1D) array of IJJs which is a stack of IJJs singled out from inside a piece of $Bi_2Sr_2CaCu_2O_{8+x}$ (BiSrCaCuO-2212, or BSCCO) single crystal, and some other circuits are integrated to it. The IJJs are quite uniform regarding the dc transport properties. Shapiro steps up to 2.5 THz and zero-crossing steps at 1.6 THz indicate that the IJJs are very sensitive to terahertz signal and the power distribution is quite even in one stack[10]. In the present work, this fabrication process is adopted and somehow modified to fabricate 2D and 3D arrays of IJJs. The basic idea is that intrinsic Josephson junctions of desirable number are located in different stacks which, in turn, are connected to each other in series. Over an area of 150 microns by 170 microns, presently we are able to fabricate 256 in-series connected stacks containing a total number of 11,000 junctions. The critical currents of the junctions scatter in a quite small region. The effects of heating and quasiparticle injection are also found remarkably reduced. At 760 GHz zero-crossing Shapiro steps are observable up to 0.75 V and 50 K.

For clarity, only the fabrication of two stacks is discussed here (Fig. 1 (a) to (c)) to emphasize the modifications from the fabrication of one stack [9]. To begin with (Fig. 1(a)), a piece of BSCCO single crystal is cleaved, fixed on Substrate 1, and plated with gold; on which a mesa with two stacks (marked with a dashed circle in the figure) is photo-lithographically patterned and fabricated. The mesa is then glued onto another substrate (Substrate 2 in Fig. 1(b)) and cleaved from the big pedestal, resulting in a Π-shaped sample about 200~300 nm thick standing on Substrate 2. After sputtering a 50 nm gold onto the sample, we put a photo resist strip on top of the sample to protect its central part; and Ar ion milling is used to etch the rest of it down to a certain depth as required by the desired junction number (from a few to a few tens) (Fig. 1(c)). Gold electrodes can be prepared in the fabrication. Note that, also for clarity, polyimide layers used for fixing the sample onto Substrate 1 (Fig. 1 (a)) or Substrate 2 (Fig. 1(b)) are not shown in the figures, nor are the gold coatings shown in Fig. 1 (a) or Fig. 1(b). While any kind of material can be used as Substrate 1, we often use Si or MgO as Substrate 2 considering their low loss at microwave frequencies for practical applications. Besides, when MgO substrate is used, we can take optical photos of the samples using not only topside illumination but also backside one.

The Π-shaped structure shown in Fig.1 is actually a "building block" in further fabrication of an array so that more junction stacks can be included. For example, if along a line we pattern a Π-shaped structure followed by an inverted Π-shaped one and again a Π-shaped one (Fig.1(d)), we end up with four stacks which are connected in series via gold on BSCCO (Au/BSCCO, the bar of the Π-shaped structure) or via BSCCO on gold (BSCCO/Au, the bar of the inverted Π-shaped structure). Using this process, we have fabricated samples with 4, 64, 128 and 256 stacks in series to each other. Usually these stacks are arranged in a plane, but not along a line, thus making the whole structure a 3-dimensional array of IJJs. Shown in Fig. 2 are the photos of the 64-stack (Figs. 2(a) and (b)) and 256-stack (Figs. 2(c)) samples taken through an optical microscope with illuminations from different angles. In Fig. 2(a) one can see, under topside illumination, bright Au/BSCCO spots and the antenna integrated to the array, while BSCCO/Au spots manifest themselves as darker ones. With MgO substrate and backside illumination, one can see in Fig. 2(b) the sample structure in detail. One of the stacks is marked by an open square, and the connecting BSCCO/Au or Au/BSCCO strips are visible and clearly distinguishable. So far we have been able to pattern up to 256 stacks in an array, which is partly shown in Fig. 2(c). Shown in Fig. 3 are the typical *I-V* curves of such a 256-stack array at 50 K measured using a bias whose maximum output is 17 volts. Due to the lack of high output bias, we are not able to trace all the resistive branches of the array (about 250 volts are needed to trace all the branches). The flatness of the *I-V* curve envelope indicates that the stacks in our array are quite homogenous, although in any particular stack the more junctions are biased at voltage states the lower the critical currents will be. We attribute the latter phenomenon to Joule heating and quasiparticle injection[3,4]. From the *I-V* curves in Fig.3, the total number of junctions in this 256-stack array is estimated to be about 11,000[11].

In our earlier work, we reported clear zero-crossing Shapiro steps in a single stack of 17 junctions with irradiation at 1.6 THz. However, the step heights were not very large and it seemed difficult to bias the stack in the step. Thus it will be very interesting to find out whether or not one can get much larger steps, thus making some practical applications possible. While attempts are made to improve the coupling at 1.6 THz, we also test the responses at a relatively low frequency of 760 GHz using the same cryostat and far infrared laser (FIR). For samples with 1 stack, 4 stacks, and 128 stacks, shown in Figs. 4(a), 4(b) and 4(c) respectively are the typical *I-V* curves under irradiation at 760 GHz. In the case of a single stack of 18 junctions (Fig. 4 (a)), we can observe 18 sharp and large zero-crossing Shapiro steps spaced at 1.57 mV, satisfying the well-known Josephson frequency-voltage relation[12]. As discussed earlier, the observed zero-crossing steps are actually the first order Shapiro step of each junction in the stack, i.e., the number of zero-crossing steps is equal to the number of junctions which are biased at their resistive states and irradiated with FIR power[10].

In the case of a 4-stack array involving about 116 junctions, we can observe zero-crossing steps at voltages up to 0.18 V, indicating that 116 junctions in series in the stack are all in their voltage states each contributing a voltage of 1.57 mV. Shown in Fig. 4 (b) are only the first 64 vertical steps with constant voltage intervals.

Shown in Fig. 4 (c) is the typical response of the 128-stack array at a relatively large scale. As marked by an arrow, the zero-crossing steps appear at a voltage as high as 0.75 V, indicating that only 480 junctions in the array are contributing; however, according to our design there are altogether 5000 junctions in this 128-stack array.

This raises a question about how many junctions in an array are able to contribute, and whether or not the corresponding Shapiro steps are zero-crossing. According to our design, there are about 5,000 and 11,000 junctions in series to each other in the 128- and 256-stack arrays, respectively. With irradiation at 760 GHz, if each junction contributes a voltage of 1.57 mV, the highest voltage will be 8 V and 17 V respectively according to the Josephson frequency-voltage relation. There are at least two reasons for the absence of zero-crossing steps at expected voltages. One is that we are not able to trace all the resistive branches of the array due to lack of large bias; in addition, biasing all junctions to their voltage states with voltage up to a few hundred volts is technically impractical. Another reason is that the coupling between the FIR signal and the stacks hasn't been optimized. Stacks close to the center of a bowtie antenna may get enough power to induce zero-crossing steps; other "distant" stacks in the array may find far less power being fed to them. The large scattering of step heights shown in Fig. 4 (c) reflects the unevenness of power distribution in different stacks, leaving much room to us for further improvements in the coupling. A multi-branch feeding structure similar to those used in niobium-based quantum voltage standards might be helpful in this regard [5,6].

For some practical applications, signal sources at millimeter or centimeter wavebands are preferred in order to achieve zero-crossing steps, mainly due to easy availability of such sources. However, the plasma frequencies of our IJJs are usually a few hundred gigahertz[10], higher than the signal frequencies at millimeter or centimeter wavebands; accordingly, their responses at millimeter or centimeter wavebands can be chaotic and no zero-crossing steps have been observed. Thus the appearance of zero-crossing Shapiro steps at terahertz reported here is of particular interest, although there is some effort to reduce the plasma frequencies of IJJs.

In summary, using a double-side fabrication method recently developed, 3-dimensional arrays of IJJs in BSCCO single crystals are realized. The measured current-voltage characteristics have shown that the junctions are quite uniform, even for the sample with 256 stacks where there are more than 10,000 junctions within an area of 170 microns by 150 microns. Zero-crossing steps are observed up to 50 K and 0.75 V, and are believed to be possible at even higher voltages, indicating a high potential for practical applications such as in quantum voltage standards. It is essential to improve the coupling between the array and the high frequency signal source, and to improve the evenness of the power distribution among each junction. From the point of view of practical applications, it is also helpful if we can reduce the plasma frequency of the IJJs and thus to justify the use of signal source at relatively low frequency.

**Acknowledgements**

This work was supported by CREST (Core Research for Evolutional Science and Technology) of Japan Science and Technology Corporation (JST) and Marubun Research Promotion Foundation, Japan, and partially carried out at the Laboratory for Electronic Intelligent Systems, Research Institute of Electrical Communication, Tohoku University, Japan.

Captions

Fig.1 (a) to (c) A schematic description of the major steps in fabricating two stacks of intrinsic Josephson junctions (IJJs) in series with each other; (d) A Π-shaped structure is used as a building block to include more stacks in an array.

Fig.2 Optical photos of a 64-stack array of IJJs on MgO substrate taken through an optical microscope with (a) topside illumination and (b) backside illumination, while shown in (c) is a photo of 256-stack array with backside illumination.

Fig.3 Typical current-voltage characteristics of a 256-stack array with a bias voltage up to 17 V.

Fig. 4. Zero-crossing Shapiro steps with irradiation at 760 GHz in (a) a single stack with 18 junctions, (b) 4 stacks with about 116 junctions, and (c) 128 stacks with about 11,000 junctions.

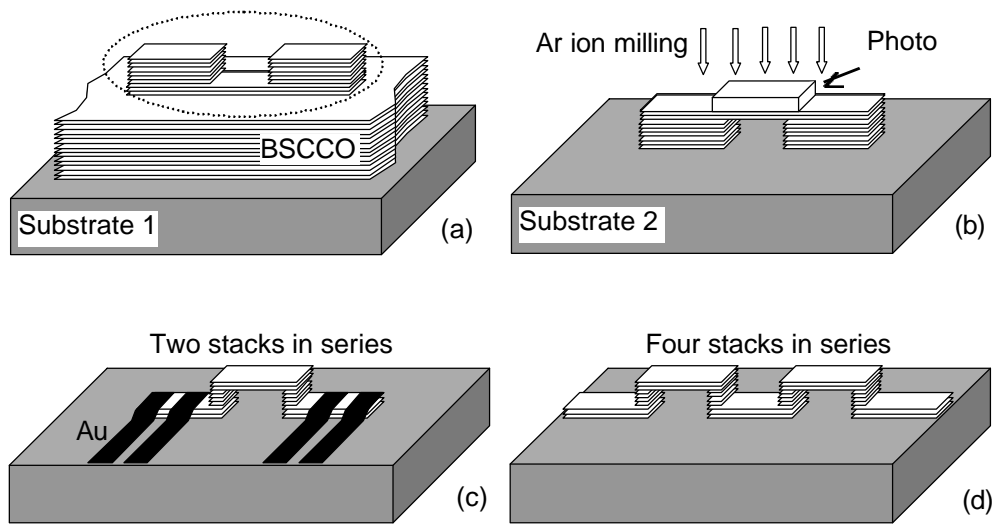

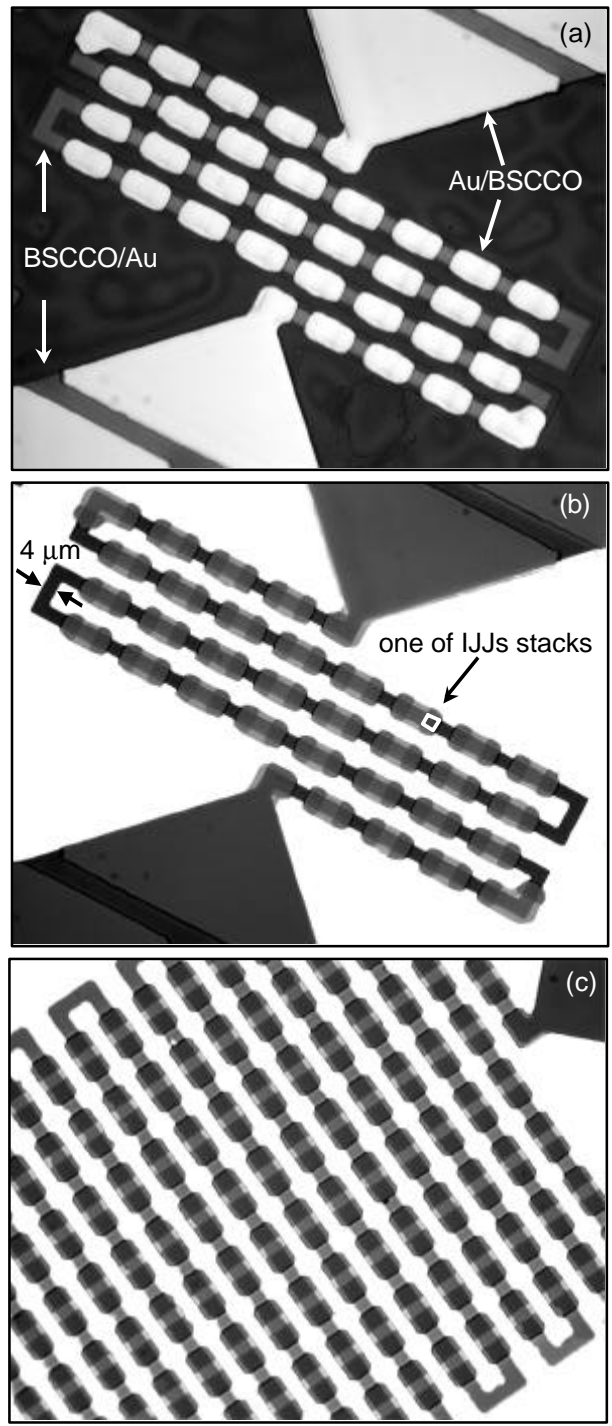

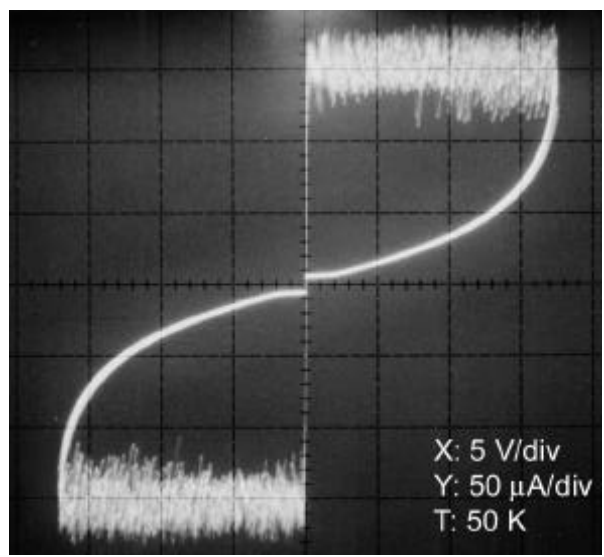

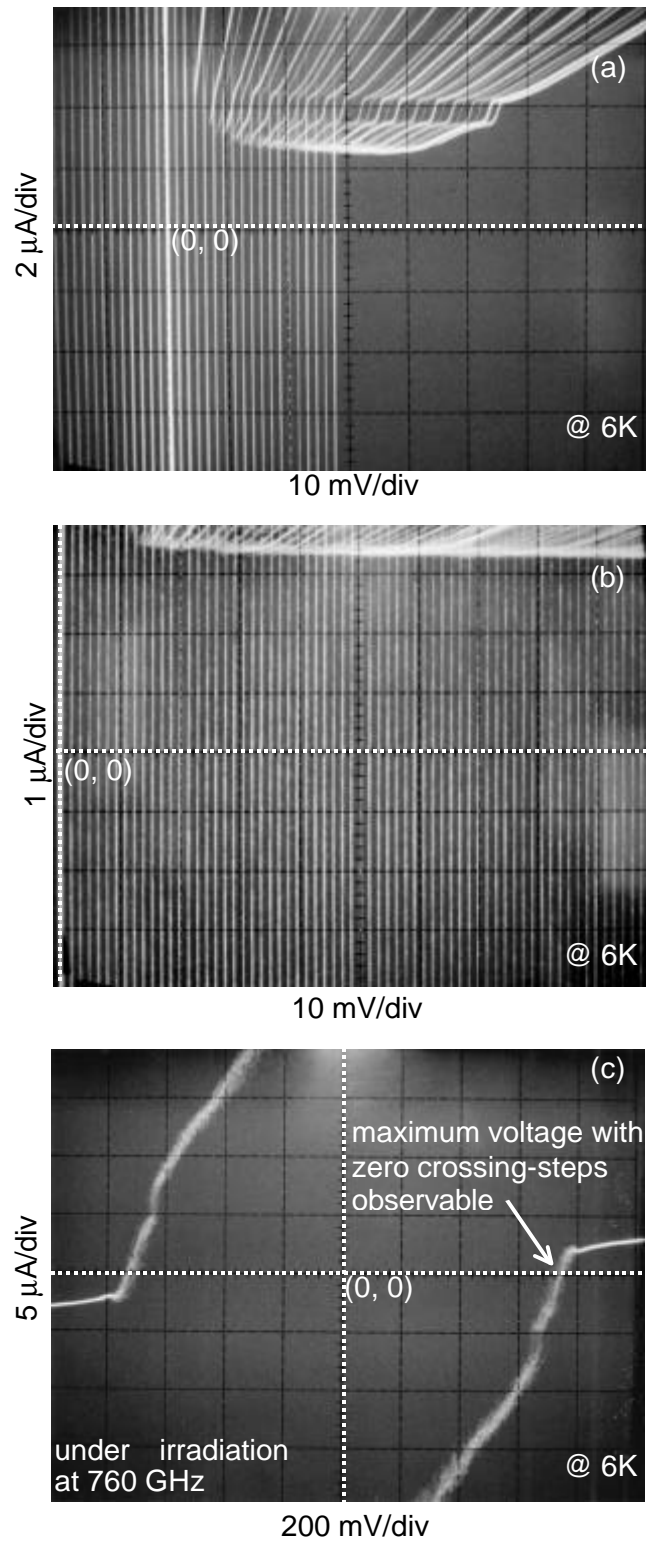

References

[1] R. Kleiner, F. Steinmeyer, G. Kunkel, and P. Müller, *Phys. Rev. Lett.* **68**, 2394 (1992).

[2] G. Oya, N. Aoyama, A. Irie, S. Kishida, and H. Tokutaka, *Jpn. J. Appl. Phys.* **31**, L829 (1992).

[3] M. Itoh, S. Karimoto, K. Namekawa, M. Suzuki, *Phys. Rev. B* **55,** R12001, 1997.

[4] A. Odagawa, M. Sakai, H. Adachi, K. Setsune, *Jpn. J. Appl. Phys.* **37,** 486 (1997).

[5] C. A. Hamilton, F. L. Lloyd, K. Ghich, and W. Goecke, IEEE Trans. Instrum. Meas. 38, 314 (1989).

[6] R. Pöpel, J. Niemeyer, R Fromknecht, W. Meier, and L. Grimm, J. Appl. Phys. 68, 4294 (1990).

[7] A. M. Klushin and H. Kohlstedt, J. Appl. Phys. 77, 441 (1995).

[8] A. M. Klushin, S. Schornstein, and H. Kohlstedt, IEEE Trans. Appl. Supercond. 7, 2423 (1997).

[9] H. B. Wang, P. H. Wu, and T. Yamashita, *Appl. Phys. Lett.* **78,** 4010 (2001)

[10] H. B. Wang, P. H. Wu, and T. Yamashita, *Phys. Rev. Lett.* **87,** 107002 (2001).

[11] H. B. Wang, K. Maeda, J. Chen, P. H. Wu, and T. Yamashita, submitted.

[12] B. D. Josephson, *Phys. Lett.*, 1, 251 (1962).